\documentclass[aps,twocolumn,superscriptaddress,nofootinbib]{revtex4-2}
\usepackage{amsmath}
\usepackage{amsfonts}
\usepackage{amssymb}
\usepackage{mathtools}
\usepackage{graphicx}
\usepackage[caption=false,position=t]{subfig}
\usepackage[colorlinks]{hyperref}
\usepackage[capitalize]{cleveref}
\usepackage{circuitikz}

\ctikzset{
	bipoles/length=0.8cm,
	bipoles/capacitor/height=.25,
	bipoles/capacitor/width=.07,
	bipoles/squid/height=.3,
	bipoles/squid/width=.3,
	inductors/scale=0.8,
	bipoles/openbarrier/gap=0}

\usepackage{yypreamble}
\newcommand{\wq}{\omega_{\rm q}}
\newcommand{\C}{\mathcal C}
\newcommand{\sz}{\hat\sigma^{\rm z}}
\newcommand{\sx}{\hat\sigma^{\rm x}}

\begin{document}

\title{Low Overhead Quantum Bus with Coupling beyond the Nearest Neighbor via Mediated Effective Capacitance}

\author{Yariv Yanay}
\email{yanay@umd.edu}
\author{Charles Tahan}
\affiliation{Laboratory for Physical Sciences, 8050 Greenmead Dr., College Park, MD 20740}

\begin{abstract}
The design of easy to operate high-fidelity two qubit gates remains an area of ongoing research. Many of the common schemes require dedicated controls lines, while others are vulnerable to issues of frequency crowding. Here, we propose a scheme for coupling a chain of transmons acting as logical qubits via a quantum bus of auxiliary qubits. The auxiliary array is made of floating transmons, and through the use of mediated interactions we generate effective capacitance between them beyond the nearest neighbor. 
Logical qubits are not directly coupled to each other, but they can be coupled by bringing them closer in frequency to the far-detuned auxiliary arrays.
This allows for tunable coupling between non-neighboring logical qubits, and for the application of entangling gates to three or more qubits at once.
\end{abstract}
\maketitle

\section{Introduction}

Recent years have seen great progress in the use of superconducting circuits for quantum computation, quantum information processing and quantum simulation applications \cite{Kjaergaard2020, Arute2019,Ma2019,Ye2019,Yan2019,Andersen2020,Braumuller2021,Tazhigulov2022}. One of the most successful architectures is the transmon qubit and its variants, which use a phase degree of freedom to store the qubit state \cite{Koch2007, Barends2013, Larsen2015}. Transmons can be connected through a simple capacitive shunt, but this coupling is always on, and leads to a small longitudinal (ZZ) coupling. A variety of proposals improve on this, including the tunable coupler with addition flux controls \cite{Geller2015, Yan2018a} or parametric driving \cite{Rigetti2005, Ashhab2007, Rigetti2010} for qubits with different frequencies. 
While these schemes have been quite successful in producing high-fidelity two qubit gates \cite{GoogleAIQuantum2020, Stehlik2021, Mitchell2021}, they are not without their flaws. Tunable couplers require an additional set of control fields, increasing the problem of magnetic flux cross-talk, while parametric driving schemes are vulnerable to issues of frequency crowding as the number of qubits grows. In addition, schemes for entangling three or more qubits are rare.

Another approach, which has not been explored to the same degree, is the use of a so-called quantum bus to connect multiple qubits, including ones that are not adjacent \cite{Aldana2011, Lu2022}. This has generally been done via a single-mode cavity, sometimes again using parametric driving approaches suffering from the same frequency crowding issues.
Here, we return to the idea of a quantum bus but take a different tack. In place of a cavity, we use an array of auxiliary qubits to mediate interaction. Our design requires no flux controls for the bus, only for the logical qubits. These can be coupled simply by bringing them near resonance with the auxiliary qubits, or entirely decoupled by tuning them away from those.

\begin{figure}[t] 
	\subfloat{\label{fig:sketchcirc}}
	\subfloat{\label{fig:sketchHfull}}
	\subfloat{\label{fig:sketchHeff}}
   \centering
\begin{circuitikz}[scale=1.7]	
	\def\dx{1}
	\def\dy{1.1}
	\def\Nq{4}
	\node at (0,\dy+0.5) {(a)};
	\foreach \y/\lab in {0/`q' array logical qubits}
	{
		\foreach \q in {1,...,\Nq}
		{
			\fill[fill=cyan,opacity=0.5,rounded corners] (\dx*\q,\y) ++(-0.25,-0.25)
				-- ++(0.2,-0.2) -- ++(0.5,0.5) -- ++(-0.4,0.4) -- ++(-0.5,-0.5) -- ++(0.2,-0.2);
			\draw (\dx*\q,\y) ++(-0.25,-0.25)
				to[C] ++(-0.2,-0.2) node[tlground]{} ++(0.2,0.2)
				to[short,*-] ++(0.15,0.15) -- ++(-0.1,0.1) to[C] ++(0.2,0.2) -- ++(0.1,-0.1)
				++ (-0.2,-0.2) -- ++(0.1,-0.1) to[squid] ++(0.2,0.2) -- ++(-0.1,0.1) to[short,-*] ++(0.15,0.15)
				to[C] ++(0.2,0.2) node[tlground,rotate=180]{}
				;
		}
		\node at (0.5*\dx-0.1,\y) [above,rotate=90,text width=1.9cm,text centered] {\lab};
	}
	\foreach \s/\lab in {+/`b' array A-B bus,-/`a' array A-A bus}
	{
		\foreach \q in {1,...,\Nq}
		{
			\fill[fill=orange,opacity=0.5,rounded corners] (\dx*\q,\s\dy) ++(-0.25,-0.25)
				-- ++(0.2,-0.2) -- ++(0.5,0.5) -- ++(-0.4,0.4) -- ++(-0.5,-0.5) -- ++(0.2,-0.2);
			\draw (\dx*\q,\s\dy) ++(-0.25,-0.25)
				to[C] ++(-0.2,-0.2) node[tlground]{} ++(0.2,0.2)
				to[short,*-] ++(0.15,0.15) -- ++(-0.1,0.1) to[C] ++(0.2,0.2) -- ++(0.1,-0.1)
				++ (-0.2,-0.2) -- ++(0.1,-0.1) to[openbarrier] ++(0.2,0.2) -- ++(-0.1,0.1) to[short,-*] ++(0.15,0.15)
				to[C] ++(0.2,0.2) node[tlground,rotate=180]{}
				;
			\draw (\dx*\q-0.25,-0.25) -- (\dx*\q-0.25,\s0.25) to[C] (\dx*\q-0.25,\s\dy-\s0.25) -- (\dx*\q-0.25,\s\dy-0.25);
			\draw (\dx*\q+0.25,+0.25) -- (\dx*\q+0.25,\s0.25) to[C] (\dx*\q+0.25,\s\dy-\s0.25) -- (\dx*\q+0.25,\s\dy+0.25);
		}
		\foreach \q in {1,...,3}
		{
			\draw (\dx*\q+\s0.25,\s\dy\s0.25) -- ++(0.25-\s0.25,0)  to[C] (\dx*\q+\dx-0.25,\s\dy-0.25);
		}
		\node at (0.5*\dx-0.1,\s\dy) [above,rotate=90,text width=1.6cm,text centered] {\lab};
	}
	\draw (1*\dx-0.25,\dy-0.25) -- ++(-\dx/2+0.25+0.05,\dx/2-0.25-0.05) to[C] ++(-0.1,0.1);
	\draw (\Nq*\dx+0.25,\dy+0.25) -- ++(\dx/2-0.25-0.05,-\dx/2+0.25+0.05) to[C] ++(0.1,-0.1);
	\draw (1*\dx-0.25,-\dy-0.25) -- ++(-\dx/2+0.25+0.05,0) to[C] ++(-0.1,0);
	\draw (\Nq*\dx-0.25,-\dy-0.25) -- ++(\dx/2+0.25-0.05,0) to[C] ++(0.1,0);
\end{circuitikz}

\begin{tikzpicture}[scale=1]
	\def\dx{0.8}
	\def\dy{0.6}
	\def\Nq{4}
	
	\foreach \q in {1,...,\Nq}
	{
		\node (A\q) at (\q*\dx,1*\dy) [circle,fill=orange!50,minimum width=12] {};
		\node (Q\q) at (\q*\dx,0*\dy) [circle,fill=cyan!50,minimum width=12] {};
		\node (B\q) at (\q*\dx,-1*\dy) [circle,fill=orange!50,minimum width=12] {};
	}
	\foreach \qa in {1,...,\Nq}
	{
		\foreach \qb in {1,...,\Nq}
		{
			\def\wC{0.5/2^abs(\qb-\qa)}
			\draw[line width=\wC] (Q\qa) to (A\qb);
			\draw[line width=\wC] (Q\qa) to (B\qb);
		}
	}
	\foreach \qa/\qb in {1/2,2/3,3/4}
	{
		\def\w{3/2^(\qb-\qa)}
		\draw[line width=\w] (A\qa) to (A\qb);
		\draw[line width=\w] (B\qa) to (B\qb);		
	}
	\foreach \qa/\qb in {1/3,2/4}
	{
		\def\w{3/2^(\qb-\qa)}
		\def\ang{30}
		\draw[line width=\w] (A\qa) to[bend left=\ang] (A\qb);
		\draw[line width=\w] (B\qa) to[bend right=\ang] (B\qb);		
	}
	\foreach \qa/\qb in {1/4}
	{
		\def\w{3/2^(\qb-\qa)}
		\def\ang{35}
		\draw[line width=\w] (A\qa) to[bend left=\ang] (A\qb);
		\draw[line width=\w] (B\qa) to[bend right=\ang] (B\qb);		
	}
	
	\def\Xeff{\Nq*\dx+1.5*\dx}
	\def\dyE{0.9*\dy}
	\foreach \ex/\on/\off/\nn/\notnn [evaluate=\ex as \y using ((2-\ex)*\dyE)] in 
		{1/{}/{1,...,4}/{}/{},
		 2/{1,3}/{2,4}/{}/{1/3},
		 3/{1,2,3}/{4}/{1/2,2/3}/{1/3}}
	{
		\draw [->](\Nq*\dx+0.3,0.1*\y) -- (\Xeff+\dx/2,\y);
		\foreach \q [evaluate=\q as \x using \Xeff+\q*\dx] in \off
			\node (Q\ex\q) at (\x,\y) [circle,fill=red!50] {};
		\foreach \q [evaluate=\q as \x using \Xeff+\q*\dx] in \on
			\node (Q\ex\q) at (\x,\y) [circle,fill=green!50] {};
		\foreach \qa/\qb in \nn
			\draw[line width=3/2] (Q\ex\qa) to (Q\ex\qb);
		\foreach \qa/\qb in \notnn
		{
			\def\w{3/2^abs(\qb-\qa)}
			\draw[line width=\w] (Q\ex\qa) to[bend left=20] (Q\ex\qb);
		}
		\draw[dashed,rounded corners] (\Xeff+\dx-0.25,\y-0.25) rectangle ++(\Nq*\dx-\dx+0.5,0.5);
	}

	\node[anchor=east] at (\dx,\dy+0.5) {(b)};
	\node[anchor=east] at (\Xeff+\dx/2,\dy+0.5) {(c)};
	\node[anchor=north] at (\dx/2+\Nq*\dx/2,-\dy-0.5) {Full Hamiltonian};
	\node[anchor=north] at (\Xeff+\Nq*\dx/2+\dx/2,-\dy-0.5) {Effective Hamiltonian};
	\node[anchor=south, fill=red!50, rounded corners] at (\Xeff+3*\dx/2,\dyE+0.4) {$\omega_{\rm q}\ll \omega_{\rm a}$};
	\node[anchor=south, fill=green!50, rounded corners] at (\Xeff+\Nq*\dx-\dx/2,\dyE+0.4) {$\omega_{\rm q}\sim \omega_{\rm a}\vphantom{\ll}$};
\end{tikzpicture}
   \caption{The mediated interaction coupling scheme. 
   \protect\subref{fig:sketchcirc} An array of tunable-frequency floating transmon logical qubits (shaded cyan) are not directly connected to each other. Instead, each qubit is linked to its counterparts in two arrays of fixed-frequency floating auxiliary transmons (shaded orange) with nearest-neighbor capacitive connections.
   Mediated interactions generate a Hamiltonian with ranged coupling, as illustrated in \protect\subref{fig:sketchHfull}, where each qubit is represented by a circle and each coupling term by a line with width proportional to its magnitude. 
   \protect\subref{fig:sketchHeff} Effective Hamiltonian of the logical qubit system. Qubits that are far detuned from the auxiliary bus (red) remain decoupled, while those tuned closer to the bus (green) gain an effective coupling to each other.
   }
   \label{fig:sketch}
\end{figure}
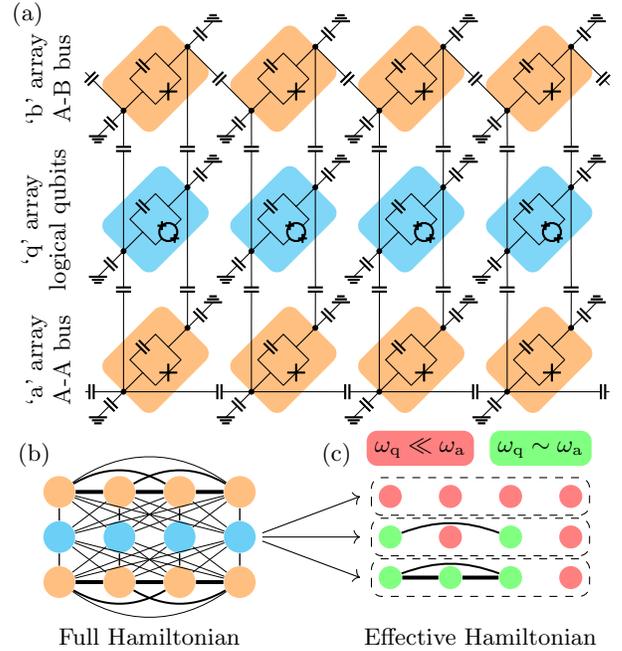

The fundamental component in our approach is the ability to generate effective capacitance beyond the nearest neighbor in a chain of so-called ``floating'' transmons, with a geometric drop-off that can be controlled by design independently of its overall strength \cite{Yanay2022}. While the method is quite versatile, the qubit coupling generated in this way is determined by the capacitance between different circuit elements, which is generally determined in fabrication and cannot be easily modified during operation. Our scheme, outlined in \cref{fig:sketch}, relegates this coupling to the auxiliary bus. The logical qubits can be set to a default coupling-off state or to a coupled regime simply by tuning their frequency. A similar scheme was recently realized experimentally \cite{Pekker2014, Scigliuzzo2022,Zhang2023}; we provide an analytical analysis of the operation mode and explore a circuit design based on the mediated interaction framework. 

A capacitively connected auxiliary array of the type shown in \cref{fig:sketch} generates coupling between the logical qubits in two ways: direct coupling, from parasitic capacitance generated via mediated interactions, and indirect coupling, via non-resonant virtual transitions in the auxiliary qubits. Crucially, we find that the use of two auxiliary buses, with opposite-sign coupling, allows for a dual cancellation effect: the parasitic capacitance between the logical qubits is completely eliminated, leaving no direct coupling; and the indirect coupling can be cancelled by tuning the logical qubit frequency to an intermediate value. Note that that two types of coupling do not cancel each other, as in some tuneable coupling elements \cite{Yan2018a}; instead, opposite-sign contributions from the two arrays cancel each one individually. Thus, for the ``off'' state, the logical qubit frequencies can be close, within few MHz, greatly alleviating frequency crowding.

The rest of the this paper is organized as follows. In \cref{sec:quant}, we briefly review circuit quantization and consider the implications for bus arrays of this form. In \cref{sec:twobus} we present our design and consider the behavior of logical qubit coupling as a function of their detuning from the auxiliary qubit arrays. In \cref{sec:implement} we consider realistic experimental parameters, and explore the effects of variance in the circuit capacitance and critical current.

\section{Overview and direct coupling cancelation\label{sec:quant}}

To explain our scheme, we begin by shortly reviewing the basics of circuit quantization. Given a circuit with some capacitive and inductive elements and $\mathcal N$ nodes, we describe it by a set of $\mathcal N$ phase differences, ${\vec \phi = \{\phi_{1},\dotsc,\phi_{\mathcal N}\}}$, and corresponding fluxes, ${\vec\Phi = \Phi_{0}\vec\phi/2\pi}$ where $\Phi_{0}$ is the magnetic flux quantum. The system Lagrangian can be written as 
\begin{equation}\begin{gathered}
\mathcal L = \half\vec{\dot\Phi}\cdot \C\cdot\vec{\dot\Phi} - \mathcal V\p{\vec\phi},
\label{eq:Lvecs}
\end{gathered}\end{equation}
where $\C$ is some \emph{capacitance matrix} defined by the capacitive elements of the circuit and $\mathcal V$ is a potential energy given by the inductive elements. The Hamiltonian is then
\begin{equation}\begin{gathered}
\mathcal H = \half\vec{q}\cdot \C^{-1}\cdot\vec{q} + \mathcal V\p{\vec\phi},
\label{eq:Hvecs}
\end{gathered}\end{equation}
where the elements of $\vec q$ are the conjugate charges of $\vec \phi$.

If the inductive elements consist of $\mathcal N$ Josesphson junctions such that $\mathcal V = \sum_{m=1}^{\mathcal N}E_{{\rm J},m}\cos\phi_{m}$, operated in the transmon regime, then the Hamiltonian can be transformed into the familiar qubit form,
\begin{equation}
\hat H \approx \sum_{m=1}^{\mathcal N}\frac{\omega_{m}}{2}\sz_{m} +\sum_{m>n}J_{mn}\sx_{m}\sx_{n},
\label{eq:Htransmon}
\end{equation}
where $\hat \gs^{\alpha}_{m}$ is the $\alpha$ Pauli matrix on qubit $m$, and the qubit frequency $\omega_{m}$ and coupling terms $J_{mn}$ are given by
\begin{equation}
\omega_{m} = \sqrt{4e^{2}\C^{-1}_{mm}E_{{\rm J},m}}, \quad 
J_{mn} = \frac{\sqrt{\omega_{m}\omega_{n}}}{2\sqrt{\C^{-1}_{mm}\C^{-1}_{nn}}}\C^{-1}_{mn}.
\label{eq:Htransmonvals}
\end{equation}
Here $e$ is the elementary charge, and we explicitly see that the inverse capacitance matrix plays the role of an inter-qubit coupling matrix.

This direct relation between the capacitance and qubit coupling becomes more subtle when there are non-qubit degrees of freedom involved. To see this, consider the situation where only some of the phases have inductive elements, $\mathcal V = \sum_{m=1}^{\mathcal N\pr}E_{{\rm J},m}\cos\phi_{m}$ for $\mathcal N\pr<\mathcal N$. We divide the vectors into those containing the the qubit degrees of freedom, $\vec\phi_{\rm q} = \{\phi_{1},\dotsc,\phi_{\mathcal N\pr}\}$, and the extraneous ones, $\vec\phi_{\rm x} = \{\phi_{\mathcal N\pr+1},\dotsc,\phi_{\mathcal N}\}$. Then we can rewrite
\begin{equation}
\mathcal L = \frac{1}{2}\mat{\vec{\dot{\Phi}_{\rm q}}\\ \vec{\dot{\Phi}_{\rm x}}}\cdot \mat{\C_{\rm qq} & \C_{\rm qx} \\ \C_{\rm xq} & \C_{\rm xx}}\cdot\mat{\vec{\dot{\Phi}_{\rm q}}\\ \vec{\dot{\Phi}_{\rm q}}} - \mathcal V\p{\vec\phi_{\rm q}}.
\end{equation}

The extraneous fluxes, $\vec{\Phi}_{\rm x}$, have no inductive component, and can be demoted into constants. However, this must be done at the Hamiltonian level, yielding an effective coupling matrix,
\begin{equation}
H = \half \vec{q}_{\rm q}\cdot \br{\C^{\rm eff}}^{-1} \cdot\vec{q}_{\rm q} +\mathcal V\p{\vec{\phi}_{q}} + O\p{\vec{q}_{\rm x}}.
\end{equation}
The direct capacitance, $\C_{\rm qq}$, is augmented by an effective portion mediated via the discarded degrees of freedom,
\begin{equation}
\C^{\rm eff} = \C_{\rm qq} - \C_{\rm qx}\cdot \br{\C_{\rm xx}}^{-1}\cdot \C_{\rm xq}.
\label{eq:ceff}
\end{equation}

\subsection*{Floating qubit auxiliary arrays}

The proposed setup relies on our previous findings regarding ``floating'' transmon qubits. A traditional transmon consists of a single pad of superconducting metal connected to ground via a Josephson junction. The phase across the junction like behaves an anharmonic oscillator, allowing it to be used as the qubit degree of freedom. A floating transmon consists of two superconducting pads connected by a Joesphson junction. Once again the qubit degree of freedom is the phase across the junction, but there is now a second, extraneous phase, from the qubit to the ground. These have been in widespread experimental use (see e.g.~\cite{Sete2021a,Stehlik2021}).

As mentioned above, because there is no inductive element the extraneous phase can be discarded, but this generates an effective capacitance.  As we have previously shown \cite{Yanay2022}, utilizing these mediated degrees of freedom one can orchestrate a geometric drop-off for this coupling strength,
\begin{equation}
\br{\C^{\rm eff}}^{-1}_{mn} = \frac{1}{C_{\rm A}}\p{\gd_{mn} + \gk \xi^{\abs{m-n}}}
\label{eq:ceffinv}
\end{equation}
for any $C_{\rm A}$ and 
\begin{equation}
0\le \xi \le 1, \qquad -\frac{1-\xi}{1+\xi} \le \gk \le \frac{1-\xi}{1+\xi}.
\end{equation}
A negative $\gk$ is achieved using the ``A-B'' configuration, shown in the top part of \cref{fig:sketchcirc}, by alternating the coupling between both sides of the transmon; a positive $\gk$ is achieved using the ``A-A'', bottom of \cref{fig:sketchcirc}, where the coupling is always on the same side.

\subsection*{Two-bus system}

The ability to vary the sign of $\gk$ is instrumental to the scheme we propose here. Consider a simpler circuit, using a single bus. The capacitance matrix of the system is given by
\begin{equation}
\C = \mat{\C_{\rm qq} & \C_{\rm qa} \\ \C_{\rm aq} & \C_{\rm aa}}
\end{equation}
where $\br{\C_{\rm qq}}_{mn}$ is the qubit-qubit capacitance matrix, $\br{\C_{\rm qa}}_{mn} = -C_{\rm qa}\gd_{mn}$ describes the capacitance between each qubit and its auxiliary correspondent, and $\C_{\rm aa}$ is the capacitance matrix of the auxiliary system with a form as in \cref{eq:ceffinv}. 
Per \cref{eq:ceff}, 
\begin{equation}\begin{gathered}
\C_{\rm qq}^{\rm eff} = \C_{\rm qq} - C_{\rm qa}^{2} \br{\C_{\rm aa}}^{-1}.
\end{gathered}\end{equation}
The qubits become directly coupled, as the effective capacitance of the qubit system inherits the longer-ranged properties of the auxiliary array. This means that we cannot turn off the coupling.

Instead, we propose the full system shown in \cref{fig:sketchcirc}. The capacitance matrix takes the form
\begin{equation}
\C = \mat{\C_{\rm qq} & \C_{\rm qa} & \C_{\rm qb} \\ \C_{\rm aq} & \C_{\rm aa} & 0  \\ \C_{\rm bq} & 0 & \C_{\rm bb}}
\end{equation}
where now $\br{\C_{\rm qb}}_{mn} = -C_{\rm qb}\gd_{mn}$. Then one finds
\begin{equation}\begin{gathered}
\C_{\rm qq}^{\rm eff} = \C_{\rm qq} - C_{\rm qa}^{2} \br{\C_{\rm aa}}^{-1}  - C_{\rm qb}^{2} \br{\C_{\rm bb}}^{-1}.
\end{gathered}\end{equation}
If we take $\C_{\rm bb}$ to have the same drop-off rate $\xi$ as $\C_{\rm aa}$ but opposite-sign $\gk$, we can set the capacitance so that the contributions from the upper and lower chain cancel out, leaving no parasitic capacitance and hence no direct coupling between the logical qubits. 

\section{Behavior of the two-bus system\label{sec:twobus}}

We fully derive the Hamiltonian of the system shown in \cref{fig:sketchcirc} in \cref{app:circderiv}, finding
\begin{subequations}\begin{gather}
\hat H \approx \hat H_{\rm q} + \hat H_{\rm a} + \hat H_{\rm b} + \hat H_{\rm qa} + \hat H_{\rm qb},
\\ \hat H_{\rm q} = \sum_{m}\frac{\omega_{\rm q}}{2}\sz_{{\rm q},m},
\\ \hat H_{\ga} = \frac{\omega_{\ga}}{2}\br{\sum_{m}\sz_{\ga,m} + \sum_{m\ne n}\frac{\gk_{\ga}\xi^{\abs{m-n}}}{1+ \gk_{\ga}}\hat\gs^{+}_{\ga,m}\hat\gs^{-}_{\ga,n}},
\\ \begin{split}\hat H_{{\rm q}\ga} & = \gve\sqrt{\omega_{\rm q}\omega_{\ga}}\abs{\gk_{\rm a}\gk_{\rm b}}^{1/4}\times
	\\ &\sum_{m}\frac{\gd_{mn} + \gk_{\ga}\xi^{\abs{m-n}}}{2\sqrt{\abs{\gk_{\ga}}}\sqrt{1+\gk_{\ga}}}\hat\gs^{+}_{{\rm q},m}\hat\gs^{-}_{{\ga},n} + \hc,
\end{split}
\end{gather}
\label{eq:Hqab}%
\end{subequations}
where $\ga=\rm a,b$. Here, $\hat\gs^{\tau}_{i,m}$ are the Pauli $\tau$ operator for qubit $m$ on the $i$ array. 
The values of the various constants are given explicitly in \cref{app:circderiv}. 

We have used the typical rotating wave approximation here, taking charge operator coupling terms to qubit ladder operators, $\hat n_{i}\hat n_{j} \to \hat\gs^{+}_{i}\hat\gs^{-}_{j} + \hat\gs^{+}_{j}\hat\gs^{-}_{i}$, and discarded the third and higher energy levels of each qubit. Generally, higher order terms of the form $\hat\gs^{\rm z}_{i}\hat\gs^{\rm z}_{j}$ also appear, but note that because of the cancelation discussed above these only connect logical qubits to \emph{auxiliary} qubits, and not to other logical qubits. As the buses are operated at their ground state, these terms can be discarded.  Finally, in our analytic calculations we take the arrays to be infinite. Finite size corrections are small as long as the array is larger than the effective coupling range $\log 1/\xi$.

Note that we have constructed the system to eliminate any direct coupling terms between the logical qubits, but the mediated interactions still generate a coupling between non-adjacent logical qubits and auxiliary qubits, with a similar geometric drop-off as in the auxiliary qubits themselves.

Next, we discuss the indirect coupling via the auxiliary arrays. We take these to be far-detuned from each other, with the tunable logical qubit frequency sitting in between. We can then neglect the bus-bus coupling, and the effective coupling decomposes into the sum of contributions from each auxiliary array, which can be calculated separately.

The system is operated with the logical qubits strongly detuned from the auxiliary arrays, ${\abs{\gve\sqrt{\omega_{\rm q}\omega_{\ga}}} \ll \abs{\omega_{\rm q} - \omega_{\ga}}}$. The auxiliary modes then generate an effective interaction between the qubits via virtual transition. For two qubits with a single common mode, this effective coupling is given by $\hat H_{\rm eff}= \frac{g^{2}}{\Delta}\p{\hat\gs^{+}_{1}\hat\gs^{-}_{2} + \hat\gs^{+}_{2}\hat\gs^{-}_{1}}$ \cite{Blais2004}. In the presence of many modes, we must first diagonalize the auxiliary system. 

\begin{figure}[thb] 
   \centering
	\includegraphics[]{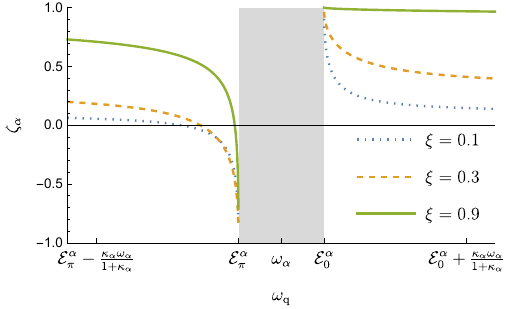}
   \caption{Behavior of the effective geometric drop-off rate, $\zeta_{\ga}$, as the logical qubit array is tuned near one of the auxiliary arrays, see \cref{eq:effpars}. 
   The grayed out area denotes the spectrum of the auxiliary array. For $\omega_{\rm q}\gg \omega_{\ga}, \omega_{\rm q}\ll \omega_{\ga}$, the effective drop-off approaches the design drop-off, $\zeta_{\ga}\to \xi$. However, as we approach the auxiliary spectrum, with its edges at $\mathcal E^{\ga}_{0},\mathcal E^{\ga}_{\pi}$, we see an enhancement, and $\zeta_{\ga}\to \pm1$, depending on the direction of approach.
   }
   \label{fig:effzeta}
\end{figure}

Because the arrays are operated at the vacuum state, it is sufficient to diagonalize the single-excitation manifold. We do this in \cref{app:Heff} and calculate effective Hamiltonian,
\begin{equation}
\hat H_{\rm q}^{\rm eff} = \sum_{m}\frac{\bar \omega_{\rm q}}{2}\sz_{{\rm q},m} + \sum_{m,n}J^{\rm eff}_{\abs{m-n}}\hat\gs^{+}_{{\rm q},m}\hat\gs^{-}_{{\rm q},n},
\label{eq:Hqeff}
\end{equation}
with the effective coupling taking the form
\begin{equation}
J^{\rm eff}_{\abs{m-n}} = J_{\rm a}\zeta_{\rm a}^{\abs{m-n}} - J_{\rm b}\zeta_{\rm b}^{\abs{m-n}},
\end{equation}
and the coupling parameters being
\begin{subequations}\begin{gather}
\bar \omega_{\rm q} = \omega_{\rm q} + \gve^{2}\omega_{\rm q}\smashoperator{\sum_{\ga={\rm a,b}}}
	 \frac{\abs{\gk_{\rm a}\gk_{\rm b}}^{1/2}}{4\abs{\gk_{\ga}}\p{1+\gk_{\ga}}}\frac{\xi}{\zeta_{\ga}}\frac{\omega_{\ga}}{\tilde\Delta^{\ga}_{\pi/2}},
\\  J_{\ga} = \frac{1}{2}\gve^{2}\abs{\gk_{\rm a}\gk_{\rm b}}^{1/2}\omega_{\rm q}
	\frac{\xi}{\zeta_{\ga}}\frac{\p{\omega_{\rm q} - \omega_{\ga}/2}^{2}}{\abs{\tilde \Delta^{\ga}_{\pi/2}}\sqrt{\Delta^{\ga}_{0}\Delta^{\ga}_{\pi}}},
	\label{eq:Ja}
\\ \zeta_{\ga} = \frac{\xi\p{\Delta^{\ga}_{\pi}+\Delta^{\ga}_{0}} + \frac{1+\xi^{2}}{2}\p{\Delta^{\ga}_{\pi}-\Delta^{\ga}_{0}}}
	{\xi\p{\Delta^{\ga}_{\pi}-\Delta^{\ga}_{0}}+\frac{1+\xi^{2}}{2}\p{\Delta^{\ga}_{\pi}+\Delta^{\ga}_{0}}}
	 \frac{\frac{1+\xi^{2}}{2}\abs{\Delta^{\ga}_{\pi/2}}}{\abs{\tilde \Delta^{\ga}_{\pi/2}}}.
\end{gather}
\label{eq:effpars}%
\end{subequations}
Here $\Delta^{\ga}_{k}$ is the effective detuning from mode $k$ of bus $\ga$, and ${\tilde \Delta^{\ga}_{\pi/2} = \Delta^{\ga}_{\pi/2}\br{\frac{1+\xi^{2}}{2} + \frac{1-\xi^{2}}{2}\sqrt{\Delta^{\ga}_{0}\Delta^{\ga}_{\pi}}/\abs{\Delta^{\ga}_{\pi/2}}}}$. See \cref{app:Heff} for details.

The form of the parameters in \cref{eq:effpars} seems at first counterintuitive. This is partly because the strength of the capacitive coupling of transmon qubits scales with the qubit frequencies, as seen in \cref{eq:Htransmonvals}. However, we note that at large detuning, $\abs{\Delta^{\ga}_{k}}\to\infty$, the effective drop-off rate approaches the engineered one $\zeta_{\ga} \approx \xi$, and the coupling tends to $J_{\rm a}\approx J_{\rm b}$. Here we see the cancellation of indirect coupling, as $J^{\rm eff}_{\abs{m-n}}$ vanishes at $\omega_{\rm a}\ll \omega_{\rm q}\ll\omega_{\rm b}$.

When $\Delta^{\ga}_{k}$ becomes relatively small, we see an enhancement in the range of the interactions. As $\Delta^{\ga}_{0}\to0$ or $\Delta^{\ga}_{\pi}\to0$, at the two edges of the auxiliary array spectrum, we find that $\zeta_{\ga} \to +1,-1$, respectively. This allows for coupling of qubits even beyond the natural range of those of the auxiliary array. This is shown in \cref{fig:effzeta}.

To avoid hybridization with the auxiliary modes, the operating regime is limited by the requirement that $\gve\sqrt{\omega_{\rm q}\omega_{\ga}} \ll \abs{\Delta^{\ga}_{0}},\abs{\Delta^{\ga}_{\pi}}$. From \cref{eq:Ja}, this implies $J_{\ga} \ll \abs{\gk_{\rm a}\gk_{\rm b}}^{1/2}\omega_{\ga}/8$.

\subsection*{Selective coupling}

Note that while our analysis considered all qubits to be set to the same frequency, the scheme can be operated selectively. In particular, consider operating the qubits in one of three modes, ${\omega_{\rm q}\in \{\omega_{\rm a+}, \omega_{\rm off},\omega_{\rm b-}\}}$, having
\begin{equation}
\omega_{\rm a}\lesssim \omega_{\rm a+} \ll \omega_{\rm off} \ll \omega_{\rm b-} \lesssim \omega_{\rm b},
\end{equation}
with $\omega_{\rm off}$ chosen as described above to eliminate the indirect coupling between the qubits, while $\omega_{\rm a+}, \omega_{\rm b-}$ are near resonant frequencies close to $\omega_{\rm a}, \omega_{\rm b}$ respectively.

Any two qubits operated at different frequencies are decoupled by virtue of being strongly detuned from each other. Within each group, the effective coupling parameters will be set according to \cref{eq:effpars}. The system's effective Hamiltonian then becomes
\begin{equation}\begin{split}
\hat H_{\rm q}^{\rm eff} & \approx \smashoperator{\sum_{m\in {\rm off}}}\frac{\bar \omega_{\rm off}}{2}\sz_{{\rm q},m}
	\\ &  + \smashoperator{\sum_{m\in {\rm a+}}}\frac{\bar \omega_{\rm a+}}{2}\sz_{{\rm q},m} + 
		\smashoperator{\sum_{m,n\in {\rm a+}}}\br{J^{\rm eff}_{\abs{m-n}}}_{\omega_{\rm a+}}\hat\gs^{+}_{{\rm q},m}\hat\gs^{-}_{{\rm q},n}
	\\ &  + \smashoperator{\sum_{m\in {\rm b-}}}\frac{\bar \omega_{\rm b-}}{2}\sz_{{\rm q},m} + 
		\smashoperator{\sum_{m,n\in {\rm b-}}}\br{J^{\rm eff}_{\abs{m-n}}}_{\omega_{\rm b-}}\hat\gs^{+}_{{\rm q},m}\hat\gs^{-}_{{\rm q},n}.
\end{split}\end{equation}

Thus, any subset of qubits can be made to interact by bringing their frequencies near resonance with one of the buses, while the remainder of the qubits idle at the \emph{off} frequency and remain decoupled. For example, in a three qubit array, we can turn on the coupling only between the first two,
\begin{equation*}\begin{gathered}
\omega_{{\rm q},1}=\omega_{\rm a+},\quad\omega_{{\rm q},2}=\omega_{\rm a+},\quad \omega_{{\rm q},3}=\omega_{\rm off},
\\ \Rightarrow \hat H_{\rm eff} \propto J^{\rm eff}_{1}\hat \gs^{+}_{1}\hat \gs^{-}_{2} + \hc,
\end{gathered}\end{equation*}
or do the same with the first and third qubits without involving the intermediate one, as in the middle panel of \cref{fig:sketchHeff},
\begin{equation*}\begin{gathered}
\omega_{{\rm q},1}=\omega_{\rm a+},\quad\omega_{{\rm q},2}=\omega_{\rm off},\quad \omega_{{\rm q},3}=\omega_{\rm a+},
\\ \Rightarrow \hat H_{\rm eff} \propto J^{\rm eff}_{2}\hat \gs^{+}_{1}\hat \gs^{-}_{3} + \hc,
\end{gathered}\end{equation*}
or bring all three near resonance and turn on all coupling terms, as in the bottom panel of \cref{fig:sketchHeff},
\begin{equation*}\begin{gathered}
\omega_{{\rm q},1}=\omega_{\rm a+},\quad\omega_{{\rm q},2}=\omega_{\rm a+},\quad \omega_{{\rm q},3}=\omega_{\rm a+},
\\ \Rightarrow \hat H_{\rm eff} \propto J^{\rm eff}_{1}\hat \gs^{+}_{1}\hat \gs^{-}_{2} + J^{\rm eff}_{1}\hat \gs^{+}_{2}\hat \gs^{-}_{3} + J^{\rm eff}_{2}\hat \gs^{+}_{1}\hat \gs^{-}_{3} + \hc
\end{gathered}\end{equation*}

\begin{figure}[btp] 
   \centering
	\includegraphics{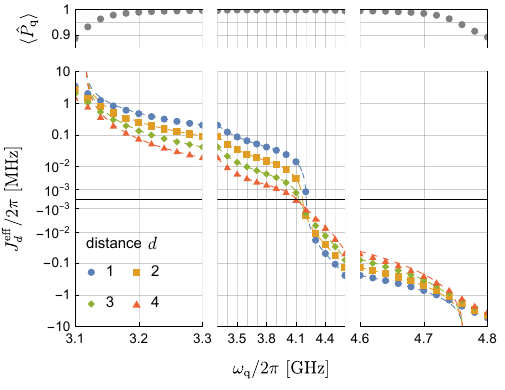}

   \caption{Behavior of the effective coupling in a our scheme. We plot the effective qubit-qubit coupling, ${J^{\rm eff}_{d} = \bvac \hat\gs^{-}_{{\rm q},n+d}\hat H^{\rm eff}_{\rm q} \hat\gs^{+}_{{\rm q},n}\vac}$, as a function of the logical qubit frequency $\omega_{\rm q}$. Different curves show the effective coupling at distance going from $d=1$ (nearest neighbors) to $d=4$. 
   As we approach the auxiliary arrays, at $\omega_{\rm a}/2\pi = 3$ GHz and $\omega_{\rm b}/2\pi = 5$ GHz, the effective interaction strength increases, with either a positive or negative sign. 
   The markers are numerically calculated, while the dashed lines are the theoretical prediction given by \cref{eq:Hqeff,eq:effpars}. 
   Above, we show the projection of the effective eigenmodes to the logical qubit manifold.
   Here we have a chain of 11 qubits with periodic boundary conditions, $\omega_{\rm a}/2\pi = 3\unit{GHz}$, $\omega_{\rm b}/2\pi = 5\unit{GHz}$, $\xi =0.3$, $\gk_{\rm a} = -\gk_{\rm b} = 0.1$, $\gve=0.01$. 
   The coupling terms are calculated by diagonalizing the modes of the Harmonic equivalent of \cref{eq:Hqab} and eliminating the auxiliary arrays.
   }
   \label{fig:numex}
\end{figure}

\section{Implementation in Superconducting Circuits\label{sec:implement}}

Having considered the fundamental physics of the scheme, we now focus on the particulars of experimental implementation. 

We begin by considering the desired coupling scales. We observe from \cref{fig:effzeta} that the effective drop-off rate $\zeta$ remains positive for any detuning $\sign\br{\Delta^{\ga}_{k}} = \sign\br{\gk_{\ga}}$, while with opposite detuning it cross from strongly negative to positive. As mentioned, $\gk_{\rm a}>0$ for the ``A-A'' configuration, and $\gk_{\rm b}<0$ for the ``A-B'' configuration. We therefore set the frequencies $\omega_{\rm a} \le \omega_{\rm q} \le \omega_{\rm b}$, taking for concreteness a target of $\omega_{\rm a}/2\pi = 3 \unit{GHz}$, $\omega_{\rm b}/2\pi = 5 \unit{GHz}$, with the resting position of the logical qubits at $\wq/2\pi = 4\unit{GHz}$.

\begin{figure}[thb] 
   \centering
	\includegraphics{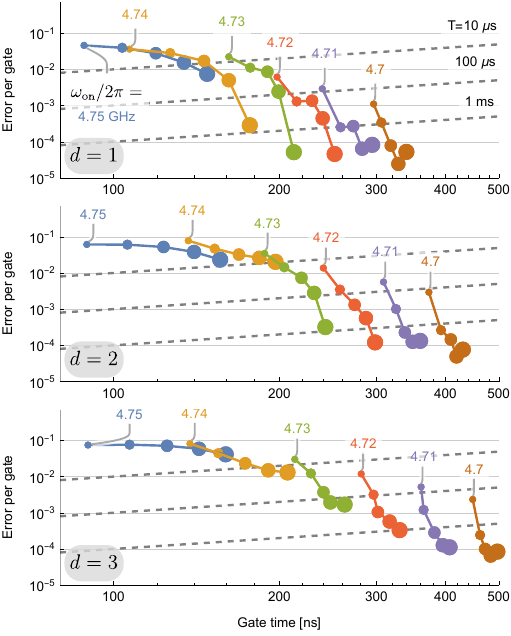}

   \caption{Gate error for a SWAP operation between qubits at distance $d=1,2,3$ (nearest, next nearest, and third-nearest neighbors), as a function of the gate operation time. 
   Here, each curve shows the gate operated at a different near-resonant qubit frequency $\omega_{\rm on}$, approaching the auxiliary array at $\omega_{\rm b}/2\pi = 5$ GHz. Different marker sizes correspond to ramping $\wq$ between the on and off frequencies over a period ranging from 10 ns (smallest) to 50 ns (largest), with on resonance time optimized for minimal gate error.
   Dashed lines show the corresponding error rate from decoherence, for qubit lifetime ranging from 10 $\rm{\mu s}$ to 1 ms.
    The calculation was done using QuTiP \cite{Johansson2013} in a 7-qubit system with the parameters as in \cref{fig:numex}, using the rotating wave Hamiltonian of \cref{eq:Hqab} but also incorporating the third level of the logical transmons.
   }
   \label{fig:gateerror}
\end{figure}

\begin{figure}[thb] 
   \centering
	\includegraphics{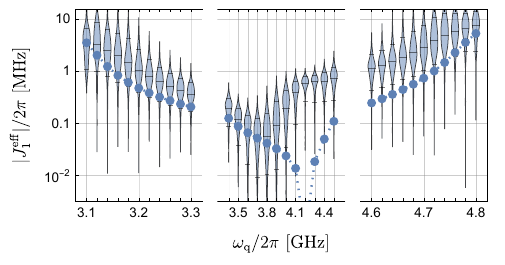}

   \caption{Behavior of the effective coupling in a two-bus system with imperfect coupling. For each value of the qubit frequency $\omega_{\rm q}$, violin plots, with lines at the 10\%, 50\% and 90\% quantiles, show the distribution of the effective nearest-neighbor logical qubit coupling across 100 realizations where all capacitances and critical currents in the circuit are normally distributed around their design values with standard deviation of 2\%. Markers, connected by dotted lines to guide the eye, show the result of the non-variance calculation. The parameters here are the same as in \cref{fig:numex}. Longer-ranged coupling, not plotted here, shows similar behavior.
   }
   \label{fig:variance}
\end{figure}

The effectiveness of the system is generally greatest for larger values of $\xi$ and $\gk_{\ga}$, allowing stronger coupling and to longer-ranged interactions. This is limited mostly by fabrication capabilities, as the ratio between the ground capacitance and coupling capacitance behave as
$C^{\ga}_{\rm G}/\C^{\ga}_{\rm c}= \p{1-\xi}^{2}\p{1+\gk_{\ga}/\frac{1-\xi}{1+\xi}}/2\xi$, see \cref{eq:Cvals}. A value of $\xi\approx 0.3$, achieved at $C^{\ga}_{\rm G}\sim\C^{\ga}_{\rm c}$, is feasible \cite{Yanay2022}, as long as $\abs{\gk_{\ga}}\ll 1$.

We consider a typical implementation of the scheme, with realistic parameters, in \cref{fig:numex}. 
We observe good agreement with \cref{eq:Hqeff,eq:effpars} in the strongly detuned regime. As the modes begin to hybridize, the effective coupling does not diverge, but we achieve an effective coupling on the order of $J^{\rm eff}/2\pi = 1-10\unit{MHz}$ out to the fourth-nearest neighbor.

To understand the full dynamics of the gates beyond the perturbative calculation, we numerically simulate the application of a SWAP gate between qubits at different distances in a 7-qubit system. The results are shown in \cref{fig:gateerror}. There is a clear tradeoff between gate fidelity and faster operation, which is achieved either through operation closer to the resonant frequencies or a faster ramp of the qubit frequencies from the idling on position. We find that to keep the error $\epsilon \lesssim 10^{-2}$, the fastest SWAP can be achieved in $t\approx 140$ ns for nearest-neighbors or $t\approx 200$ ns for the next- and third-nearest neighbors, corresponding to a coupling strength of $J^{\rm eff}\approx 1-2$ MHz.

Finally, we consider variation in the critical currents and capacitances of the circuit. With modern techniques these variances can be reduced to about 1-3\% \cite{Osman2021}. We numerically calculate the effective Hamiltonian for the same implementation as above with all parameters normally varying with a standard deviation of 2\%. The results for the nearest-neighbor coupling are shown in \cref{fig:variance}. Surprisingly, we observe that the effective coupling tends to be stronger than it is for the design with precise parameters.

The increased inter-qubit coupling comes from an increase in the effective logical-auxiliary capacitance (and so coupling), $\C_{\rm qa}$, $\C_{\rm qb}$. These in turn are mediated via the discarded ``+'' degrees of freedom of these auxiliary arrays. The scheme as designed and shown in \cref{fig:sketchcirc} uses symmetric coupling to avoid these terms, but the variation in capacitance values breaks this symmetry. 

The same effect also increases the coupling in the strongly detuned regime, which we use to decouple the logical qubits from each other when idling. This can be seen in the middle panel of \cref{fig:variance}. However, we find that there is still a range where these effective couplings can be reduced to $\abs{J^{\rm eff}}/2\pi \ll 1\unit{MHz}$, so that by slightly detuning the logical qubits from each other we can keep them decoupled.

\section{Outlook\label{sec:outlook}}

We've shown here how a circuit based on the mediated capacitance principle can be used as a quantum bus for a set of tunable logical qubits. The inter-qubit coupling can be tuned up to several megahertz, either positive or negative, or turned off entirely.

There are multiple potential uses for this scheme. First, it allows for a new kind of architecture with on-off couplers that has a simple structure, does not suffer from frequency crowding, and requires just a single flux control for each logical qubit. This substantially reduces issues of cross-talk and may enable achieving higher fidelities in a scalable manner. Second, the architecture enables coupling beyond the nearest neighbor, allowing for versatility and a reduced number of operations in various quantum algorithms. 

Finally, the ability to entangle multiple qubits at once in a controllable way, opens the path to a whole class of quantum gates that has so far not been explored. As the controls are limited to one tuneable parameter per qubit, the scheme does not allow the generation of arbitrary multi-qubit gates. However, the ability to generate coupling with geometric drop-off between any subset of the qubit, as well as tune the interaction to either positive or negative signs, adds a powerful tool to the multi-qubit toolbox.

The scheme we present here is the most straightforward implementation of the concept, but there are many other architectures that could be explored. We have found that the strength of the coupling is increased by adding a degree of asymmetry to the logical-bus coupling, and it is possible this could be used to generate a more optimal circuit if fabrication allows for more precise specification of the capacitance. We have also seen that the same coupling form can be generated in a two-dimensional grid \cite{Yanay2022}, suggesting that an even more versatile configuration could be created, perhaps with the use of integrated three-dimensional structures \cite{Rosenberg2017}.

\clearpage
\appendix
\widetext
\section{Derivation of the System Hamiltonian\label{app:circderiv}}

\begin{figure}[h] 
   \centering
\begin{circuitikz}[scale=3]	
	\def\dx{1}
	\def\dy{1.2}
	\def\Nq{4}
	\foreach \y/\lab/\tag in {0/`q' array logical qubits/q}
	{
		\foreach \q in {1,...,\Nq}
		{
			\fill[fill=cyan,opacity=0.5,rounded corners] (\dx*\q,\y) ++(-0.25,-0.25)
				-- ++(0.2,-0.2) -- ++(0.5,0.5) -- ++(-0.4,0.4) -- ++(-0.5,-0.5) -- ++(0.2,-0.2);
			\draw (\dx*\q,\y) ++(-0.25,-0.25)
				to[C,l_=$C^{\rm \tag}_{\rm G}$] ++(-0.2,-0.2) node[tlground]{} ++(0.2,0.2)
				to[short,*-] ++(0.15,0.15) -- ++(-0.1,0.1) to[C=$C^{\rm \tag}_{\rm Q}$] ++(0.2,0.2) -- ++(0.1,-0.1)
				++ (-0.2,-0.2) -- ++(0.1,-0.1) to[squid,l_=$E^{\rm q}_{J}$] ++(0.2,0.2) -- ++(-0.1,0.1) to[short,-*] ++(0.15,0.15)
				to[C,l_=$C^{\rm \tag}_{\rm G}$] ++(0.2,0.2) node[tlground,rotate=180]{}
				;
		}
		\node at (0.5*\dx-0.1,\y) [above,rotate=90,text width=1.9cm,text centered] {\lab};
	}
	\foreach \s/\lab/\tag in {+/`b' array A-B bus/b}
	{
		\foreach \q in {1,...,\Nq}
		{
			\fill[fill=orange,opacity=0.5,rounded corners] (\dx*\q,\s\dy) ++(-0.25,-0.25)
				-- ++(0.2,-0.2) -- ++(0.5,0.5) -- ++(-0.4,0.4) -- ++(-0.5,-0.5) -- ++(0.2,-0.2);
			\draw (\dx*\q,\s\dy) ++(-0.25,-0.25)
				to[C,l_=$C^{\rm \tag}_{\rm G}$] ++(-0.2,-0.2) node[tlground]{} ++(0.2,0.2)
				to[short,*-] ++(0.15,0.15) -- ++(-0.1,0.1) to[C=$C^{\rm \tag}_{\rm Q}$] ++(0.2,0.2) -- ++(0.1,-0.1)
				++ (-0.2,-0.2) -- ++(0.1,-0.1) to[openbarrier,l_=$E^{\rm q}_{J}$] ++(0.2,0.2) -- ++(-0.1,0.1) to[short,-*] ++(0.15,0.15)
				to[C,l=$C^{\rm \tag}_{\rm G}$] ++(0.2,0.2) node[tlground,rotate=180]{}
				;
			\draw (\dx*\q-0.25,-0.25) -- (\dx*\q-0.25,\s0.25) to[C=$C^{\rm q}_{\rm \tag}$] (\dx*\q-0.25,\s\dy-\s0.25) -- (\dx*\q-0.25,\s\dy-0.25);
			\draw (\dx*\q+0.25,+0.25) -- (\dx*\q+0.25,\s0.25) to[C=$C^{\rm q}_{\rm \tag}$] (\dx*\q+0.25,\s\dy-\s0.25) -- (\dx*\q+0.25,\s\dy+0.25);
		}
		\foreach \q in {1,...,3}
		{
			\draw (\dx*\q+\s0.25,\s\dy\s0.25) -- ++(0.25-\s0.25,0)  to[C=$C^{\rm \tag}_{\rm c}$] (\dx*\q+\dx-0.25,\s\dy-0.25);
		}
		\node at (0.5*\dx-0.1,\s\dy) [above,rotate=90,text width=1.6cm,text centered] {\lab};
	}
	\foreach \s/\lab/\tag in {-/`a' array A-A bus/a}
	{
		\foreach \q in {1,...,\Nq}
		{
			\fill[fill=orange,opacity=0.5,rounded corners] (\dx*\q,\s\dy) ++(-0.25,-0.25)
				-- ++(0.2,-0.2) -- ++(0.5,0.5) -- ++(-0.4,0.4) -- ++(-0.5,-0.5) -- ++(0.2,-0.2);
			\draw (\dx*\q,\s\dy) ++(-0.25,-0.25)
				to[C,l=$C^{\rm \tag}_{\rm G}$] ++(-0.2,-0.2) node[tlground]{} ++(0.2,0.2)
				to[short,*-] ++(0.15,0.15) -- ++(-0.1,0.1) to[C=$C^{\rm \tag}_{\rm Q}$] ++(0.2,0.2) -- ++(0.1,-0.1)
				++ (-0.2,-0.2) -- ++(0.1,-0.1) to[openbarrier,l_=$E^{\rm q}_{J}$] ++(0.2,0.2) -- ++(-0.1,0.1) to[short,-*] ++(0.15,0.15)
				to[C,l_=$C^{\rm \tag}_{\rm G}$] ++(0.2,0.2) node[tlground,rotate=180]{}
				;
			\draw (\dx*\q-0.25,-0.25) -- (\dx*\q-0.25,\s0.25) to[C=$C^{\rm q}_{\rm \tag}$] (\dx*\q-0.25,\s\dy-\s0.25) -- (\dx*\q-0.25,\s\dy-0.25);
			\draw (\dx*\q+0.25,+0.25) -- (\dx*\q+0.25,\s0.25) to[C=$C^{\rm q}_{\rm \tag}$] (\dx*\q+0.25,\s\dy-\s0.25) -- (\dx*\q+0.25,\s\dy+0.25);
		}
		\foreach \q in {1,...,3}
		{
			\draw (\dx*\q+\s0.25,\s\dy\s0.25) -- ++(0.25-\s0.25,0)  to[C=$C^{\rm \tag}_{\rm c}$] (\dx*\q+\dx-0.25,\s\dy-0.25);
		}
		\node at (0.5*\dx-0.1,\s\dy) [above,rotate=90,text width=1.6cm,text centered] {\lab};
	}
	\draw (1*\dx-0.25,\dy-0.25) -- ++(-\dx/2+0.25+0.05,\dx/2-0.25-0.05) to[C] ++(-0.1,0.1);
	\draw (\Nq*\dx+0.25,\dy+0.25) -- ++(\dx/2-0.25-0.05,-\dx/2+0.25+0.05) to[C] ++(0.1,-0.1);
	\draw (1*\dx-0.25,-\dy-0.25) -- ++(-\dx/2+0.25+0.05,0) to[C] ++(-0.1,0);
	\draw (\Nq*\dx-0.25,-\dy-0.25) -- ++(\dx/2+0.25-0.05,0) to[C] ++(0.1,0);
\end{circuitikz}
   \caption{Full circuit diagram for the system sketched out in \cref{fig:sketch}.
   }
   \label{fig:circuitcalouts}
\end{figure}
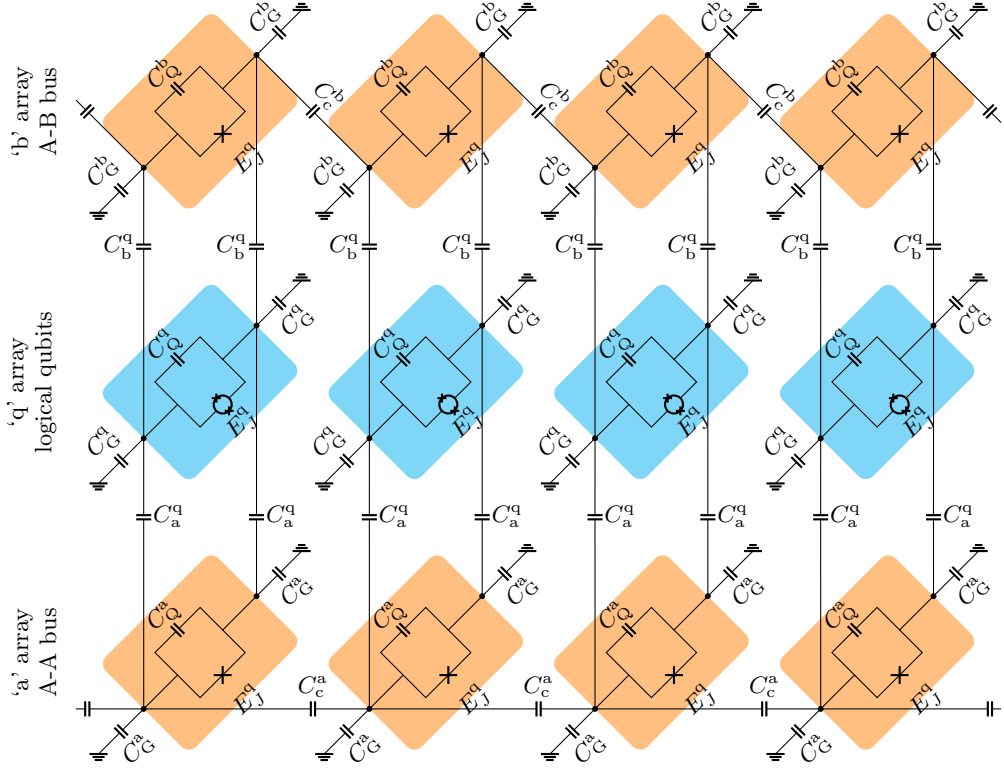

Here, we explicitly derive the system Hamiltonian for the two-bus circuit, drawn explicitly in \cref{fig:circuitcalouts}.

The Lagrangian given by
\begin{equation}\begin{split}
\mathcal L & = \sum_{\substack{x=\\\rm q,a,b}}\sum_{m=1}^{L} \frac{C^{x}_{\rm Q}}{2} \p{\dot\Phi^{x\nearrow}_{m} - \dot\Phi^{x\swarrow}_{m}}^{2} 
		+ \frac{C^{x}_{\rm G}}{2} \p{(\dot\Phi^{x\swarrow}_{m})^{2} + (\dot\Phi^{x\nearrow}_{m})^{2}} + E^{x}_{{\rm J}m}\cos\p{\phi^{x\nearrow}_{m} - \phi^{x\swarrow}_{m}}
\\ & + \sum_{\substack{\ga=\\{\rm a},b}}\sum_{m=1}^{L} \frac{C^{{\rm q}}_{\ga}}{2} \br{\p{\dot\Phi^{\ga\nearrow}_{m} - \dot\Phi^{{\rm q}\nearrow}_{m}}^{2} 
	+ \p{\dot\Phi^{\ga\swarrow}_{m} - \dot\Phi^{{\rm q}\swarrow}_{m}}^{2} }
+ \sum_{m=1}^{L-1} \frac{C^{{\rm a}}_{\rm c}}{2} \p{\dot\Phi^{{\rm a}\swarrow}_{m+1} - \dot\Phi^{{\rm a}\swarrow}_{m}}^{2}
	+ \frac{C^{{\rm b}}_{\rm c}}{2} \p{\dot\Phi^{{\rm b}\swarrow}_{m+1} - \dot\Phi^{{\rm b}\nearrow}_{m}}^{2}.
\end{split}\end{equation}
Substituting $\Phi^{x}_{m\pm} = \Phi^{x\nearrow}_{m} \pm \Phi^{x\swarrow}_{m}$ we have
\begin{equation}\begin{split}
\mathcal L & = \frac{1}{2}\dot{\vec \Phi}\cdot \C \cdot \dot{\vec \Phi} + \sum_{\substack{x=\\\rm q,a,b}}\sum_{m=1}^{L} E^{x}_{{\rm J}m}\cos\phi^{x-}_{m}
\end{split}\end{equation}
where
\begin{equation}
\C = \mat{\C_{\rm qq} & \C_{\rm qa} & \C_{\rm qb} \\ \C_{\rm aq} & \C_{\rm aa} & 0 \\ \C_{\rm bq} & 0 & \C_{\rm bb}},
\qquad
\C_{xy} = \mat{\C_{xy}^{--} & \C_{xy}^{-+} \\ \C_{xy}^{+-} & \C_{xy}^{++}},
\qquad \vec\Phi = \mat{\dot{\vec{\Phi}}^{{\rm q}-} \\ \dot{\vec{\Phi}}^{{\rm q}+} \\ \dot{\vec{\Phi}}^{{\rm a}-} \\ \dot{\vec{\Phi}}^{{\rm a}+} \\ \dot{\vec{\Phi}}^{{\rm b}+} \\ \dot{\vec{\Phi}}^{{\rm b}-}}.
\end{equation}
The capacitance matrices are
\begin{subequations}\begin{gather}
\br{\C_{\rm qq}^{--}}_{mn} = \p{C^{\rm q}_{\rm Q} + \frac{C^{\rm q}_{\rm G}}{2} + \frac{C^{\rm q}_{\rm a}}{2} + \frac{C^{\rm q}_{\rm b}}{2}}\gd_{mn}
\\ \br{\C_{\rm qq}^{++}}_{mn} = \p{\frac{C^{\rm q}_{\rm G}}{2} + \frac{C^{\rm q}_{\rm a}}{2} + \frac{C^{\rm q}_{\rm b}}{2}}\gd_{mn}
\\ \br{\C_{{\rm q}x}^{--}}_{mn} = \br{\C_{{\rm q}x}^{++}}_{mn} = -\frac{C^{\rm q}_{x}}{2}\gd_{mn}
\\ \br{\C_{\rm aa}^{--}}_{mn} = \p{C^{\rm a}_{\rm Q} + \frac{C^{\rm a}_{\rm G}}{2} + \frac{C^{\rm q}_{\rm a}}{2}}\gd_{mn} + \frac{C^{\rm a}_{\rm c}}{4}\br{2\gd_{mn} - \gd_{m+1,n} - \gd_{m-1,n}}
\\ \br{\C_{\rm aa}^{++}}_{mn} = \p{\frac{C^{\rm a}_{\rm G}}{2} + \frac{C^{\rm q}_{\rm a}}{2}}\gd_{mn} + \frac{C^{\rm a}_{\rm c}}{4}\br{2\gd_{mn} - \gd_{m+1,n} - \gd_{m-1,n}}
\\ \br{\C_{\rm aa}^{+-}}_{mn} = -\frac{C^{\rm a}_{\rm c}}{4}\br{2\gd_{mn} - \gd_{m+1,n} - \gd_{m-1,n}}
\\ \br{\C_{\rm bb}^{--}}_{mn} = \p{C^{\rm b}_{\rm Q} + \frac{C^{\rm b}_{\rm G}}{2} + \frac{C^{\rm q}_{\rm b}}{2}}\gd_{mn} + \frac{C^{\rm b}_{\rm c}}{4}\br{2\gd_{mn} + \gd_{m+1,n} + \gd_{m-1,n}}
\\ \br{\C_{\rm bb}^{++}}_{mn} = {\frac{C^{\rm b}_{\rm G}}{2} + \frac{C^{\rm q}_{\rm b}}{2}}\gd_{mn} + \frac{C^{\rm b}_{\rm c}}{4}\br{2\gd_{mn} - \gd_{m+1,n} - \gd_{m-1,n}}
\\ \br{\C_{\rm bb}^{+-}}_{mn} = \frac{C^{\rm b}_{\rm c}}{4}\br{\gd_{m+1,n} - \gd_{m-1,n}}.
\end{gather}\end{subequations}
From symmetry, $\br{\C^{\gs\tau}_{xy}}_{mn} = \br{\C^{\tau\gs}_{yx}}_{nm}$.

Per \cref{eq:Hvecs}, the Hamiltonian is determined by the inverse matrix. This can be written as 
\begin{equation}\begin{split}
\C^{-1}& = \mat{\br{\C_{\rm qq}^{\rm eff}}^{-1} & -\br{\C_{\rm qq}^{\rm eff}}^{-1}\C_{\rm qa}\br{\C_{\rm aa}}^{-1} & -\br{\C_{\rm qq}^{\rm eff}}^{-1}\C_{\rm qb}\br{\C_{\rm bb}}^{-1}
	\\  -\br{\C_{\rm aa}}^{-1}\C_{\rm aq}\br{\C_{\rm qq}^{\rm eff}}^{-1} & \br{\C_{\rm \bar q \bar q}^{\rm eff}}^{-1}_{\rm aa} & \br{\C_{\rm \bar q \bar q}^{\rm eff}}^{-1}_{\rm ab}
	\\  -\br{\C_{\rm bb}}^{-1}\C_{\rm bq}\br{\C_{\rm qq}^{\rm eff}}^{-1} & \br{\C_{\rm \bar q \bar q}^{\rm eff}}^{-1}_{\rm ba} & \br{\C_{\rm \bar q \bar q}^{\rm eff}}^{-1}_{\rm bb}
}
\end{split}\end{equation}
with
\begin{subequations}\begin{gather}
\C_{\rm qq}^{\rm eff} = \C_{\rm qq} -  \C_{\rm qa}\br{\C_{\rm aa}}^{-1}\C_{\rm aq} - \C_{\rm qb}\br{\C_{\rm bb}}^{-1}\C_{\rm bq},
\\ \C_{\rm \bar q \bar q}^{\rm eff} = {\mat{ \C_{\rm aa} & 0 \\ 0 & \C_{\rm bb}} - \mat{\C_{\rm aq}\br{\C_{\rm qq}}^{-1}\C_{\rm qa} & \C_{\rm aq}\br{\C_{\rm qq}}^{-1}\C_{\rm qb} 
	\\ \C_{\rm bq}\br{\C_{\rm qq}}^{-1}\C_{\rm qa} & \C_{\rm bq}\br{\C_{\rm qq}}^{-1}\C_{\rm qb}}}.
\end{gather}\end{subequations}

If the coupling elements $C^{\rm q}_{\rm b},C^{\rm q}_{\rm a}\ll C^{x}_{\rm Q},C^{x}_{\rm G}$, to second order in $\C_{{\rm q}x}$,
\begin{subequations}\begin{gather}
\br{\C_{\rm qq}^{\rm eff}}^{-1} \approx \br{\C_{\rm qq}}^{-1} + \br{\C_{\rm qq}}^{-1}\C_{\rm qa}\br{\C_{\rm aa}}^{-1}\C_{\rm aq}\br{\C_{\rm qq}}^{-1} 
	+ \br{\C_{\rm qq}}^{-1}\C_{\rm qb}\br{\C_{\rm bb}}^{-1}\C_{\rm bq}\br{\C_{\rm qq}}^{-1},
\\ \br{\C_{\rm \bar q \bar q}^{\rm eff}}^{-1} \approx \mat{ \br{\C_{\rm aa}}^{-1} & 0 \\ 0 & \br{\C_{\rm bb}}^{-1}} + 
	\mat{\br{\C_{\rm aa}}^{-1}\C_{\rm aq}\br{\C_{\rm qq}}^{-1}\C_{\rm qa}\br{\C_{\rm aa}}^{-1} 
	& \br{\C_{\rm aa}}^{-1}\C_{\rm aq}\br{\C_{\rm qq}}^{-1}\C_{\rm qb}\br{\C_{\rm bb}}^{-1} 
	\\ \br{\C_{\rm bb}}^{-1}\C_{\rm bq}\br{\C_{\rm qq}}^{-1}\C_{\rm qa}\br{\C_{\rm aa}}^{-1} 
	& \br{\C_{\rm bb}}^{-1}\C_{\rm bq}\br{\C_{\rm qq}}^{-1}\C_{\rm qb}\br{\C_{\rm bb}}^{-1}},
\\ \C^{-1}_{{\rm q}x} \approx -\br{\C_{\rm qq}}^{-1}\C_{{\rm q}x}\br{\C_{\rm xx}}^{-1}.
\end{gather}
\label{eq:Cineffappx}
\end{subequations}

We can now discard the extraneous degrees of freedom, to remain with the Hamiltonian
\begin{equation}\begin{split}
\mathcal H = \frac{1}{2}\mat{\vec{q}^{\rm q -} \\ \vec{q}^{\rm a-} \\ \vec{q}^{\rm b-}}\cdot \br{\C^{-1}}^{--} \cdot \mat{\vec{q}^{\rm q -} \\ \vec{q}^{\rm a-} \\ \vec{q}^{\rm b-}} - \sum_{\substack{x=\\\rm q,a,b}}\sum_{m=1}^{L} E^{{\rm q}}_{{\rm J}m}\cos\phi^{{\rm q}-}_{m}.
\end{split}\end{equation}
Using the diagonal form of $\C_{\rm qq}, \C_{{\rm q}x}$, we find immediately
\begin{subequations}\begin{gather}
\br{\C_{\rm qq}^{\rm eff}}^{-1}_{--} \approx \br{\C_{\rm qq}^{--}}^{-1} + \p{\frac{C^{\rm q}_{\rm a}}{2\bar C_{\rm q}}}^{2}\br{\C_{\rm aa}}^{-1}_{--}
	+ \p{\frac{C^{\rm q}_{\rm b}}{2\bar C_{\rm q}}}^{2}\br{\C_{\rm bb}}^{-1}_{--},
\\ \br{\C_{\rm \bar q \bar q}^{\rm eff}}^{-1}_{--} \approx \mat{ \br{\C_{\rm aa}}^{-1}_{--} & 0 \\ 0 & \br{\C_{\rm bb}}^{-1}_{--}} + 
	\mat{\frac{\p{C^{\rm q}_{\rm a}}^{2}}{4\bar C_{\rm q}}\br{\br{\C_{\rm aa}}^{-1}\br{\C_{\rm aa}}^{-1}}_{--}
		& \frac{C^{\rm q}_{\rm a}C^{\rm q}_{\rm b}}{4\bar C_{\rm q}}\br{\br{\C_{\rm aa}}^{-1}\br{\C_{\rm bb}}^{-1}}_{--}
	\\  \frac{C^{\rm q}_{\rm a}C^{\rm q}_{\rm b}}{4\bar C_{\rm q}}\br{\br{\C_{\rm bb}}^{-1}\br{\C_{\rm aa}}^{-1}}_{--}
		 & \frac{\p{C^{\rm q}_{\rm b}}^{2}}{4\bar C_{\rm q}}\br{\br{\C_{\rm bb}}^{-1}\br{\C_{\rm bb}}^{-1}}_{--}
		 },
\\ \C^{-1}_{{\rm q}x} \approx \frac{C^{\rm q}_{x}}{2\bar C_{\rm q}}\br{\C_{xx}}^{-1}_{--}.
\end{gather}
\end{subequations}
 where $\bar C_{\rm q} = {C^{\rm q}_{\rm Q} + \frac{C^{\rm q}_{\rm G}}{2} + \frac{C^{\rm q}_{\rm a}}{2} + \frac{C^{\rm q}_{\rm b}}{2}}$.
 
We design the auxiliary arrays to have specific properties. By taking 
\begin{subequations}\begin{gather}
C^{\rm a}_{\rm Q}  = \frac{1 - \gk_{\rm a}/\gk_{\rm max}}{1 + \gk_{\rm a}/\gk_{\rm max}}\bar C_{\rm a} - \frac{C^{\rm q}_{\rm a}}{2},
\\ C^{\rm a}_{\rm G} = \frac{2\gk_{\rm a}/\gk_{\rm max}}{1 + \gk_{\rm a}/\gk_{\rm max}}\bar C_{\rm a},
\\ C^{\rm a}_{\rm c} = \frac{4\xi}{\p{1-\xi}^{2}}\frac{\gk_{\rm a}/\gk_{\rm max}}{\p{1 + \gk_{\rm a}/\gk_{\rm max}}^{2}}\bar C_{\rm a},
\\ C^{\rm b}_{\rm Q}  = \bar C_{\rm b} - \frac{C^{\rm q}_{\rm b}}{2},
\\ C^{\rm b}_{\rm G} = \frac{2\gk_{\rm max}\abs{\gk_{\rm b}}}{1-\gk_{\rm max}\abs{\gk_{\rm b}}}\bar C_{\rm b},
\\ C^{\rm b}_{\rm c} = \frac{4\xi}{\p{1-\xi}^{2}}\frac{\abs{\gk_{\rm b}}\gk_{\rm max}}{\p{1-\gk_{\rm max}\abs{\gk_{\rm b}}}\p{1-\abs{\gk_{\rm b}}/\gk_{\rm max}}}\bar C_{\rm b},
\end{gather}
\label{eq:Cvals}\end{subequations}
where $\gk_{\rm max} = \frac{1-\xi}{1+\xi}$, and $-\gk_{\rm max}< \gk_{\rm b}<0<\gk_{\rm a}<\gk_{\rm max}$, we ensure for $\ga={\rm a,b}$ \cite{Yanay2022}
\begin{subequations}\begin{gather}
\br{\br{\C_{\ga\ga}}^{-1}}^{--}_{mn} = \frac{1}{\bar C_{\ga}}\br{\gd_{mn} + \gk_{\ga}\xi^{\abs{m-n}}}.
\end{gather}\end{subequations}

Further enforcing
\begin{equation}
\frac{\p{C^{\rm q}_{\rm a}}^{2}}{\bar C_{\rm a}}\gk_{\rm a} = \frac{\p{C^{\rm q}_{\rm b}}^{2}}{\bar C_{\rm b}}\abs{\gk_{\rm b}},
\end{equation}
we find
\begin{subequations}\begin{gather}
\br{\br{\C^{-1}}^{--}_{\rm qq}}_{mn} = \frac{1}{\bar C_{\rm q}}\br{1 
	+ \gve^{2}\frac{\gk_{\rm a} + \abs{\gk_{\rm b}}}{\sqrt{\abs{\gk_{\rm a}\gk_{\rm b}}}}}\gd_{mn} + O\p{\gve}^{3}
\\ \br{\br{\C^{-1}}^{--}_{{\rm q}\ga}}_{mn} = \gve\frac{\abs{\gk_{\rm a}\gk_{\rm b}}^{1/4}}{\sqrt{\abs{\gk_{\ga}}\bar C_{\rm q}\bar C_{\ga}}}
	\br{\gd_{mn} + \gk_{x}\xi^{\abs{m-n}}} + O\p{\gve}^{3},
\\ \br{\br{\C^{-1}}^{--}_{\ga\ga}}_{mn} = \frac{1}{\bar C_{\ga}}\br{\gd_{mn} + \gk_{\ga}\xi^{\abs{m-n}}} + O\p{\gve}^{2},
\\ \br{\br{\C^{-1}}^{--}_{\rm ab}}_{mn} =  O\p{\gve}^{2},
\end{gather}\end{subequations}
with
\begin{equation}
\gve^{2} = \frac{C^{\rm q}_{\rm a}C^{\rm q}_{\rm b}}{4\sqrt{\bar C_{\rm a}\bar C_{\rm b}}{\bar C_{\rm q}}}\ll 1.
\end{equation}

Because we are interested in the dynamics of the logical qubits, we can ignore the corrections to the auxiliary arrays, we will drop the correction to the a,b portions of the inverse capacitance matrix. We can now rewrite the Hamiltonian using \cref{eq:Htransmon,eq:Htransmonvals} as
\begin{subequations}\begin{gather}
\hat H \approx \hat H_{\rm q} + \hat H_{\rm a} + \hat H_{\rm b} + \hat H_{\rm qa} + \hat H_{\rm qb},
\\ \hat H_{\rm q} = \sum_{m}\frac{\omega_{\rm q}}{2}\hat \gs^{z}_{{\rm q},m},
\\ \hat H_{\rm a} = \sum_{m}\frac{\omega_{\rm a}}{2}\hat \gs^{z}_{{\rm a},m} 
	+ \frac{\gk_{\rm a}\omega_{\rm a}}{2\p{1+\gk_{\rm a}}}\sum_{m\ne n}\xi^{\abs{m-n}}\hat \gs^{+}_{{\rm a},m}\hat \gs^{-}_{{\rm a},n}
	\label{eq:Ha}
\\ \hat H_{\rm b} = \sum_{m}\frac{\omega_{\rm b}}{2}\hat \gs^{z}_{{\rm b},m} 
	+ \frac{\gk_{\rm b}\omega_{\rm b}}{2\p{1+\gk_{\rm b}}}\sum_{m\ne n}\xi^{\abs{m-n}}\hat \gs^{+}_{{\rm b},m}\hat \gs^{-}_{{\rm b},n}
	\label{eq:Hb}
\\ \hat H_{\rm qa} = \gve\sqrt{\omega_{\rm q}\omega_{\rm a}} \frac{\abs{\gk_{\rm b}/\gk_{\rm a}}^{1/4}}{2\sqrt{1+\gk_{\rm a}}}
	\sum_{m,n}\p{\gd_{mn} + \gk_{\rm a}\xi^{\abs{m-n}}}\hat\gs^{+}_{{\rm q},m}\hat\gs^{-}_{{\rm a},n} + \hc
\\ \hat H_{\rm qb} = \gve\sqrt{\omega_{\rm q}\omega_{\rm b}} \frac{\abs{\gk_{\rm a}/\gk_{\rm b}}^{1/4}}{2\sqrt{1+\gk_{\rm b}}}
	\sum_{m,n}\p{\gd_{mn} + \gk_{\rm b}\xi^{\abs{m-n}}}\hat\gs^{+}_{{\rm q},m}\hat\gs^{-}_{{\rm b},n} + \hc
\end{gather}
\label{eq:Hall}
\end{subequations}
where $\hat \gs^{\tau}_{x,m}$ is the $\tau$ Pauli operator on the $m$ site of the $\ga$ array.

\section{Effective Hamiltonian\label{app:Heff}}

We derive the effective logical qubit Hamiltonian here. First, we use
\begin{equation}
\hat \gs^{-}_{\ga,k} = \frac{1}{\sqrt{2\pi}}\sum_{m}e^{ikm}\hat \gs^{-}_{\ga,m}, \qquad \hat \gs^{-}_{\ga,m} = \frac{1}{\sqrt{2\pi}}\int_{-\pi}^{\pi}\dd k e^{-ikm}\hat \gs^{-}_{\ga,k},
\end{equation}
to diagonalize the single-excitation space of the auxiliary arrays,
\begin{equation}
\hat H_{\ga} = \omega_{\ga}\int_{-\pi}^{\pi}\dd k\p{1+E^{\ga}_{k}}\hat\gs^{+}_{\ga,k}\hat\gs^{-}_{\ga,k},
\end{equation}
where 
\begin{equation}
E^{\ga}_{k} = \frac{\gk_{\ga}}{1+\gk_{\ga}}\frac{\xi\p{\cos k - \xi}}{1 + \xi^{2} - 2\xi\cos k}.
\end{equation}

The interaction terms in this basis are given by
\begin{equation}\begin{split}
\hat H_{{\rm q}\ga} & = \gve\sqrt{\omega_{\rm q}\omega_{\ga}} \frac{\abs{\gk_{\rm a}\gk_{\rm b}}^{1/4}}{2\sqrt{\abs{\gk_{\ga}}}\sqrt{1+\gk_{\ga}}}
	\sum_{m,n}\int_{-\pi}^{\pi}\dd k\frac{e^{-ikn}}{\sqrt{2\pi}}\p{\gd_{mn} + \gk_{\ga}\xi^{\abs{m-n}}}\hat\gs^{+}_{{\rm q},m}\hat\gs^{-}_{{\ga},k} + \hc
\\ & = \gve\sqrt{\omega_{\rm q}\omega_{\ga}} \frac{\abs{\gk_{\rm a}\gk_{\rm b}}^{1/4}}{\sqrt{\abs{\gk_{\ga}}}}\sqrt{1+\gk_{\ga}}
	\sum_{m}\int_{-\pi}^{\pi}\dd k\frac{e^{-ikm}}{\sqrt{2\pi}}\p{\frac{1}{2} + E^{\ga}_{k}}\hat\gs^{+}_{{\rm q},m}\hat\gs^{-}_{{\ga},k} + \hc
\end{split}\end{equation}
and so we can write
\begin{equation}\begin{split}
\hat H_{\rm q}^{\rm eff} & = \hat H_{\rm q} + \sum_{m,n}\hat\gs^{+}_{{\rm q},m}\hat\gs^{-}_{{\rm q},n}\smashoperator{\sum_{\ga={\rm a,b}}}\gve^{2}\omega_{\rm q}\omega_{\ga}
	\frac{\abs{\gk_{\rm b}\gk_{\rm a}}^{1/2}}{\abs{\gk_{\ga}}}\p{1+\gk_{\ga}}
	\int_{-\pi}^{\pi}\dd k \frac{e^{ik\p{n-m}}}{2\pi}\frac{\p{\half+E^{\ga}_{k}}^{2}}{\omega_{\rm q} - \omega_{\ga}\p{1 + E^{\ga}_{k}}}
\\ & = \hat H_{\rm q} + \sum_{m,n}\hat\gs^{+}_{{\rm q},m}\hat\gs^{-}_{{\rm q},n}\smashoperator{\sum_{\ga={\rm a,b}}}\frac{1}{2}\gve^{2}\omega_{\rm q}
	\frac{\abs{\gk_{\rm b}\gk_{\rm a}}^{1/2}}{\abs{\gk_{\ga}}}\times
	\\ & \qquad\qquad \br{\frac{1}{2\p{1+\gk_{\ga}}}\frac{\xi}{\zeta_{\ga}}\frac{\omega_{\ga}}{\tilde\Delta^{\ga}_{\pi/2}}\gd_{mn}
	+ \gk_{\ga}\frac{\xi}{\zeta_{\ga}}\frac{\p{\omega_{\rm q} - \omega_{\ga}/2}^{2}}{\abs{\tilde \Delta^{\ga}_{\pi/2}}\sqrt{\Delta^{\ga}_{0}\Delta^{\ga}_{\pi}}}\zeta_{\ga}^{\abs{m-n}}
	- \gk_{a}\xi^{\abs{m-n}}}.
\label{eq:Hqeffapp}
\end{split}\end{equation}
where
\begin{subequations}\begin{gather}
\Delta^{\ga}_{k} = \omega_{\rm q} - \omega_{\ga}\p{1 + E^{\ga}_{k}},
\\ \tilde \Delta^{\ga}_{\pi/2} = \Delta^{\ga}_{\pi/2}\br{\frac{1+\xi^{2}}{2} + \frac{1-\xi^{2}}{2}\sqrt{\Delta^{\ga}_{0}\Delta^{\ga}_{\pi}}/\abs{\Delta^{\ga}_{\pi/2}}},
\\ \zeta_{\ga} = \frac{\xi\p{\Delta^{\ga}_{\pi}+\Delta^{\ga}_{0}} + \frac{1+\xi^{2}}{2}\p{\Delta^{\ga}_{\pi}-\Delta^{\ga}_{0}}}
	{\xi\p{\Delta^{\ga}_{\pi}-\Delta^{\ga}_{0}}+\frac{1+\xi^{2}}{2}\p{\Delta^{\ga}_{\pi}+\Delta^{\ga}_{0}}}
	 \frac{\frac{1+\xi^{2}}{2}\abs{\Delta^{\ga}_{\pi/2}}}{\abs{\tilde \Delta^{\ga}_{\pi/2}}}.
\end{gather}\end{subequations}

Note that the last term in the brackets in \cref{eq:Hqeffapp} is independent of $\ga$ except for the sign of $\gk_{\ga}$; as we set up $\gk_{\rm b}<0<\gk_{\rm a}$, the two contributions cancel out. However, note also that for large detuning $\omega_{\rm q}\to \infty$, we observe that $\zeta_{\ga}\to \xi$ and that the two last term in the brackets tend to cancel out, leaving the leading term of order $\omega_{\ga}/\omega_{\rm q}$ as expected.

\bibliography{}

\end{document}